\documentclass[aps,prl,reprint, preprintnumbers,amsmath,amssymb,superscriptaddress]{revtex4-1}

\usepackage{graphicx}
\usepackage{dcolumn}
\usepackage{bm}
\usepackage{hyperref}
\usepackage[utf8x]{inputenc} 
\usepackage{textcomp} 
\usepackage{comment} 
\usepackage{bookmark}

\usepackage{enumerate}

\newcommand{\EE}{\ensuremath{\mathbf{E}}}
\newcommand{\DD}{\ensuremath{\mathbf{D}}}
\newcommand{\BB}{\ensuremath{\mathbf{B}}}
\newcommand{\HH}{\ensuremath{\mathbf{H}}}

\newcommand{\X}{\ensuremath{\hat{\chi}}}
\newcommand{\Xt}{\ensuremath{\hat{\chi}^{T}}}

\newcommand{\chixy}{\ensuremath{\chi_{xy}} }
\newcommand{\chiyx}{\ensuremath{\chi_{yx}} }

\newcommand{\kk}{\ensuremath{\mathbf{k}}}
\newcommand{\ee}{\ensuremath{\mathbf{e}}}

\usepackage{cancel}

\begin{document}



\title{Does the Feigel effect break the first law?}

\author{Ottavio A. Croze}
\affiliation{ Cavendish Laboratory, University of Cambridge, Cambridge, CB3 0HE, United Kindom}


\date{\today}

\begin{abstract}
A recent theory posits that the quantum vacuum can transfer momentum to magnetoelectric media (the Feigel effect). Several related theories of vacuum momentum transfer to media have since been proposed. Neither these, nor the original theory have been observed experimentally, the existence a Feigel effect remaining highly contentious. Some investigations predict a measurable effect; others conclude vacuum momentum transfer to magnetoelectric media is not physically possible. Most analyses, including the original Fiegel theory, do not model experimentally realistic geometries and boundary conditions. I recently provided an alternative derivation of Feigel theory applied to realistic geometries, proposing experimental tests. I show here that in such geometries the existence of a steady Feigel effect (time-independent magnetoelectric susceptibilities) is equivalent to a violation of the first law of thermodynamics. A steady Feigel effect should not exist, as confirmed by a semi-classical quantum electrodynamic analysis. However, unsteady Feigel effects compatible with QED and thermodynamics are possible, and two such new effects are proposed.
\end{abstract}
 
\pacs{Valid PACS appear here}
\keywords{Feigel effect, quantum vacuum, thermodynamics, magnetoelectric, strong magnetic fields, vacuum radiometer}

\maketitle


Empty space is not empty at all: it teems with vacuum fluctuations. These can be thought as virtual photons with energy $\hbar \omega_\kk/2$ and momentum $\hbar \kk/2$, where $\omega_\kk$ is the angular frequency of mode with wave vector $\kk$ and $\hbar$ is Planck's angular constant. Initially it was doubted that vacuum fluctuations should have physically measurable consequences. However, vacuum effects now rest on experimentally firm ground, with the most paradigmatic examples being the Lamb shift and the Casimir effect \cite{Milonni94}. The latter, only relatively recently measured with sufficient accuracy \cite{Lamoreaux97, MohideenAnushree98}, is a particularly spectacular macroscopic manifestation of vacuum fluctuations able to drive parallel metal plates together. This can be understood in terms of vacuum momentum transfer \cite{Milonnietal88}. In the Casimir effect this momentum transfer is (statistically) symmetric. In contrast, a recent untested theory by Feigel posits that the vacuum can transfer momentum to magnetoelectric materials asymmetrically \cite{Feigel04}. This is possible because magnetoelectric molecular structure breaks the temporal and spatial symmetries of electromagnetic modes, leading to optical anisotropy and allowing asymmetric momentum transfer.

The Feigel effect is a very interesting idea and has stimulated several recent theoretical investigations \cite{vanTiggelenetal06, Shenetal06, BirkelandBrevik07, ObukhovHehl08, KawkavanTiggelen10, Croze12, RikkenvanTiggelen11, SilveirinhaMaslovski12}. The original proposal by Feigel amounts to considering a magnetoelectric (ME) liquid that fills all of space, away from boundaries. The magnetoelectric response of the fluid is induced by applying strong perpendicular electric and magnetic fields, which also need to permeate all space to avoid introducing field gradients and boundaries. Though the unbounded scenario is not experimentally realistic, some theoretical analyses have considered it concluding that the Feigel effect should vanishes when the QED theory is properly regularised \cite{vanTiggelenetal05, vanTiggelenetal06}. In alternative to effectively magnetoelectric liquids, one can instead consider solids whose magnetoelectricity can be induced by application of cooling and weak magnetic fields \cite{Jungetal04}. Obukhov \& Hehl \cite{ObukhovHehl08} considered these, evaluating the net force on an infinite magnetoelectric slab of finite thickness semi-classically. They predict a nonzero force for light on a magnetoelectric slab (real photons, e.g. due to counter propagating laser beams hitting the slab from both sides), but calculate that this force on slabs in a vacuum should vanish. Other theoretical proposals consider magnetoelectric media sandwiched between Casimir-style parallel metal plates \cite{BirkelandBrevik07, vanTiggelenetal06, SilveirinhaMaslovski12}. This allows a regularisation of the theory without introducing a cut-off as in the original Feigel theory, but predictions that have been made for momenta using these theories remain unmeasurably small \cite{vanTiggelenetal06}. Only two investigations have considered realistically measureable Feigel effects. The first is my own analysis of a realistic bounded steady Feigel effect, where I provide an alternative derivation Feigel's result, and a new expression for the vacuum stress on an magnetoelectric liquid, used to make experimentally observable prediction \cite{Croze12}. The second is the microscopic theory by Kawka and vanTiggelen \cite{KawkavanTiggelen10} who formulated a non relativistic QED theory of harmonic oscillator in crossed electric and magnetic fields, predicting finite momentum transfer. The only experimental test of a Feigel effect is by \cite{RikkenvanTiggelen11} who considered gases of atoms with magnetoelectric responses caused by oscillating perpendicular electric and magnetic fields, as in the Feigel proposal. Rikken \& van Tiggelen \cite{RikkenvanTiggelen11} rule out my correction to the transient effect originally predicted by Feigel, and state that their measurements are not sufficiently sensitive to test their own prediction \cite{KawkavanTiggelen10}: momentum transfer is dominated by the classical Abraham force contribution. No attempts to experimentally observe macroscopic Feigel effects, as I have proposed \cite{Croze12}, have yet been made. 

In this paper I show that, if a realistic Feigel effect exists for steady magnetoelectric response, the first law of thermodynamics will be violated. I then amend the QED analysis of the steady effect (which only considered half the proposed experimental geometry) \cite{Croze12}, using a semiclassical approach. This revised analysis shows that there can be no net vacuum momentum contribution in a steady Feigel effect. I will finally briefly propose new dynamical and Casimir-Feigel steady effects compatible with conservation of energy.

{\it Thermodynamic considerations.} In \cite{Croze12} I considered two experimental reasonable scenarios to test the Feigel effect: i) an organometallic dielectric liquid rendered magnetoelectric by the application of large perpendicular electric and magnetic fields; ii) a vacuum radiometer with paddles made of a magnetoelectric solid, such as a polar ferrimagnet. For case (i) I corrected minor inaccuracies in the derivation by Feigel of the momentum density in the ME fluid caused by the anisotropic propagation of vacuum modes. One can then derive, either semiclassically or through a kinetic theory argument, a new expression for the stress these modes cause on the fluid in the magnetoelectric (ME) region: $T_{\rm vac}\approx \Delta \chi \hbar c/\lambda_c^4$, where $\Delta\chi\equiv\chixy-\chiyx$ is the difference between the nonzero components of the magnetoelectric susceptibility tensor, $\hbar$ is Planck's angular constant, $c$ is the speed of light in vacuo, and $\lambda_c$ is the cut-off wavelength. This is the wavelength below which vacuum modes does not experience a ME medium, in analogy with the similar cut-off for light \cite{Feigel04, Croze12}. I used this new vacuum stress in the Navier-Stokes equation to evaluate the maximum speed and corresponding flow rate of organometallic fluid flow in a tube, a short portion of which is exposed to strong orthogonal electric and magnetic fields inducing ME response, see \cite{Croze12} for details. Considering steady fields, i.e. $d \X/dt=0$, where $\X$ is the ME susceptibility tensor, and ignoring gradient effects at the ME region boundaries, I obtained a vacuum flow speed $U_{\rm vac}=T_0 a^2/(4 \eta L)\approx 100\mu$m/s (where $a, L$ are the tube radius and length, respectively, and $\eta$ dynamic viscosity of fluid) and corresponding flow rate $\Phi_{\rm vac}=\pi a^2 U_{\rm vac}/2$ driven by a vacuum pressure of $T_{\rm vac}=0.03$ Pa. I based these estimates on reasonable experimental parameters for an organometallic fluid (again, see \cite{Croze12} for details) considering a tube with $a=1$mm and $L=2$m. Arranging this tube into a closed loop it is clear that, with $\EE$ and $\BB$ fields steady, a non-zero Feigel effect in the ME region implies the vacuum can continuously drive flow in a circle with flow rate $\Phi_{\rm vac}$. This flow will dissipate energy at a rate $\epsilon=T_{\rm vac} \Phi_{\rm vac} \sim 1$nW. Where is this energy coming from? The steady fields making the fluid magnetoelectrc cannot be putting energy into the ME material (charges can do no work in steady fields), with the implication that either the vacuum does not actually transfer momentum to the ME medium or energy (albeit in tiny amounts) can be extracted from the vacuum without other energy input. This is a blatant violation of conservation of energy the first law of thermodynamics. Similar considerations apply to the vacuum radiometer. The rotation of its magnetoelectric vanes caused by the vacuum would be dissipated by friction at the pivot at a rate $\sim \gamma \omega_{\rm vac}^2$: the vacuum would be generating temperature differences with no energy input. Something is clearly wrong in my QED analysis in \cite{Croze12}, the organometallic liquid should not flow and the radiometer should not turn (I do no wish to question conservation of energy!). The analysis, however, demonstrates quite clearly how a steady Feigel effect should not exist purely on thermodynamic grounds (one cannot extract free momentum or energy from the vacuum). For comparison, let us consider using the established Casimir effect to make a `vacuum engine'. In the Casimir scenario, the vacuum potential energy difference in the vacuum energy between the plates and outside (due to the change in the mode spectrum by the boundaries) is converted into kinetic energy of the plates until they touch and stop moving. An equal amount of work needs to be performed to move the plates apart: the first law is safe. 

{\it Revised quantum electrodynamic analysis.} The problem with my analysis in \cite{Croze12} is that it only addressed half the system. Let us follow my treatment and consider a tube (lenght $L$, radius $a$) with open ends at atmospheric pressure. The region of the tube where the fields act is denoted as region 1. In this region the strong crossed fields induce magnetoelectric (ME) susceptibilities $\chi_{ij}$, just as light is observed to induce them in organometallic liquids \cite{RothRikken02}. The portion of the tube to the right of the ME region is denoted as region 2. The situation is depicted in figure \ref{magnetflow} below. 
\begin{figure}[tbph]
\centering
\includegraphics[width=0.7\linewidth]{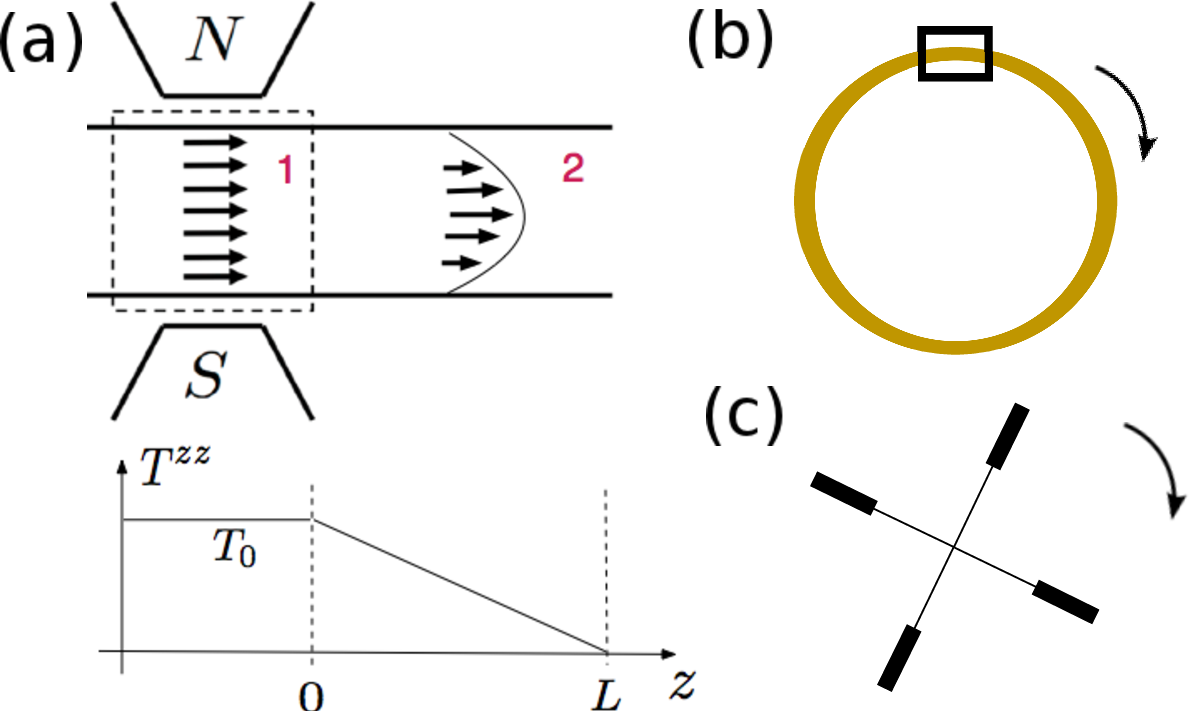}
\caption{(a) an organometallic ME fluid in crossed fields contained in a tube, as envisaged by \cite{Croze12}. The tube is divided into two regions: region 1 where the fields act (so that the fluid behaves like a magnetoelectric); and region 2 where there are no fields (and no additional stresses). (b) If the tube is a closed loop, a non-zero steady Feigel effect implies vacuum momentum could be used to generate heat without performing work; (c) similarly the vacuum momentum transfer on a ME slab would cause vacuum radiometer to turn and the dissipation at the pivot would generate useable hear. These are blatant violations of the first law of thermodynamics.}\label{magnetflow}
\end{figure}
We evaluate the electromagnetic force density in regions 1 and 2. In general this is given by (Schwinger, 1998):
\begin{equation}\label{fEM}
{\bf f}= \frac{\partial{\bf g} }{\partial t}-\nabla\cdot {\bf T}
\end{equation}
where {\bf g} is the EM momentum density and {\bf T} is the EM stress tensor.  In region 1, the fields cause the dielectric fluid to behave like a ME material with constitutive equations (Gaussian units throughout):
\begin{eqnarray}\label{Con1}
&&\DD=\epsilon \EE + \X \HH,\\
&&\BB=\mu \HH + \Xt \EE,\label{Con2}
\end{eqnarray}
where the $\X$ is the magnetoelectric susceptibility tensor, defined, in matrix form, by:
\begin{equation}\label{MESuscTensor0}
\X\equiv \left(
                           \begin{array}{ccc}
                             0 & \chixy & 0 \\
                             \chiyx & 0 & 0 \\
                             0 & 0 & 0 \\
                           \end{array}
                           \right)
\end{equation}
Following \cite{Croze12}, we evaluate `pseudo-momentum' density ${\bf g}=(\DD\times\BB-\EE\times\HH)/(4\pi c)$ which, using (\ref{Con1}), (\ref{Con2}), can be written as:
\begin{equation}\label{MElagrangian}
{\bf g}=\frac{1}{4\pi c} \left[\left( \epsilon  -\frac{1}{\mu}\right)
\EE\times\BB+\frac{1}{\mu} \EE\times(\Xt \EE) + \frac{1}{\mu}
(\X\BB)\times\BB\right]
\end{equation}
The field modes for electromagnetic waves in a magnetoelectric are given by (taking the optical axis along the z-direction, $\ee_{3}$),
\begin{eqnarray}\label{MEmodes00_1}
  &[\EE_{\pm\kk1}, \BB_{\pm\kk1}]=E_{\pm\kk1}[\ee_{1}, n_{\pm\kk1}\ee_{2}]\\\label{MEmodes00_2}
  &[\EE_{\pm\kk2}, \BB_{\pm\kk2}]=E_{\pm\kk2}[\ee_{2}, -n_{\pm\kk2}\ee_{1}]
\end{eqnarray}
where 
\begin{equation}\label{mode}
E_{\pm\kk\lambda}=E_{0 k}\cos{(k_\lambda z-\omega t)}
\end{equation}
for each polarisation $\lambda=1,2$, and where
\begin{equation*}
n_{\pm\kk,1}=\pm n_0+\chixy,\,\,\,\,\,\,\,n_{\pm\kk,2}=\pm n_0-\chiyx
\end{equation*}
Evaluating (\ref{MElagrangian}) for the counterpropagating EM modes in the z-direction using (\ref{MEmodes00_1}) and (\ref{MEmodes00_2}) 
 (see \cite{Feigel04, Croze12}) the time-averaged momentum density per wavevector $k$ can be evaluated as
\begin{equation}\label{MomDens}
\overline{{\bf g}_k}=\Delta\chi \frac{1}{c}\frac{\epsilon E^2_{0 k}}{4\pi};\mbox{~~~~~}
\overline{\frac{\partial{\bf g}_k }{\partial t}}=0
\end{equation}
where $\Delta \chi=\chixy-\chiyx$ and the over-line denotes a time-average over a period $\tau$: $\overline{{\bf A}(t)}=(1/\tau)\int_0^\tau {\bf A} dt$ for a general tensor ${\bf A}$. The term $\overline{\partial{\bf g}_k/\partial t}=0$, because, when substituting modes (\ref{mode}) into $\partial{\bf g}_{k\lambda}/\partial t$, all terms multiplied by factors $\sim\sin{[2(k_\lambda z-\omega t)]}$, averaging to zero, vanish. This leaves only terms $\propto\partial\chi_{ij}/\partial t$, but, for steady state external fields, the $\chi_{ij}$ are constant, so this contribution also vanishes. Note, however that, for unsteady fields discussed later, $\chi_{ij}= \chi_{ij}(t)$, so $\overline{\partial{\bf g}_k/\partial t}$ is in general non-zero.

One can similarly evaluate the EM stress tensor. For field modes propagating along $z$: $\nabla\cdot{\bf T}=\partial_z T^{zz}$. We thus need only calculate $T^{zz}$ (noting that the Abraham and Minkoswki expressions coincide for this term). This component is given by 
\begin{equation}\label{Tensor}
T^{zz}=-\frac{1}{4\pi c} \left[\frac{1}{2}\epsilon {\bf E}^2-\frac{1}{2\mu}{\bf B}^2\right].
\end{equation}
To calculate the mode contribution of modes we evaluate the net stress: 
\begin{equation}\label{NetTensor}
T^{zz}_k=\sum_\lambda (T^{zz}_{+\kk \lambda}-T^{zz}_{-\kk \lambda}),
\end{equation}
thus, substituting modes (\ref{MEmodes00_1}) and (\ref{MEmodes00_2}) into (\ref{Tensor}) and (\ref{NetTensor}), we obtain
\begin{equation}\label{MomDens}
\overline{T^{zz}_k}=\Delta\chi\frac{\epsilon E^2_{0 k}}{4\pi};\mbox{~~~~~}
\overline{\frac{\partial T^{zz}_k }{\partial z}}\left(=\frac{\partial \overline{T^{zz}_k }}{\partial z}\right)=0
\end{equation}
We see that the time-averaged stress is constant. Quantising EM energy density and summing over all modes leads to the vacuum momentum density $\langle 0|{\bf g}|0\rangle={\bf g_0}$ and stress $\langle 0|T^{zz}|0\rangle=T_0=g_0/c$.  It is this stress that could give rive to a Feigel effect, let us evaluate the resulting vacuum force balance. In region 1 the mean force density vanishes for steady magnetoelectrics:
\begin{equation}\label{MomDensReg1}
\overline{{\bf f_1}}= \overline{\frac{\partial{\bf g}_1 }{\partial t}}-\nabla\cdot \overline{{\bf T}_1}=0
\end{equation}
In region 2 there are no external fields to cause ME anisotropy so $\overline{{\bf f}_2}=0$. However, there are additional forces acting on the 12 boundary. Due to the ME anisotropy of region 1, we have a non-zero stress at the boundary given by $T^{zz}(z=0)=T_0$. Far from the boundary no stress acts, so $T^{zz}(z=L)=0$, where $L$ is the length of the tube. Assuming organometallic fluids are Newtonian, we have a constant stress gradient in the fluid in the tube: $\partial_z T^{zz}=C$, where $C$ is a constant. Integrating between $z=0$ and $z=L$, we find $C=T_0/L$. Thus
\begin{equation}\label{fEM2}
\overline{{\bf f}_{12}}=\frac{\partial}{\partial z} T^{zz} \kk=\frac{T_0}{L}\kk.
\end{equation}
the net force density between regions 1 and 2 is $f_{\rm vac}=\overline{\cancel{{\bf f}_1}}+\cancel{\overline{{\bf f}_2}}+\overline{{\bf f}_{12}}=\frac{T_0}{L}\kk$, and it appears that the vacuum should  indeed exert a force on the ME medium. 
This analysis, however, neglects the force on the material through the 21 boundary. By symmetry this given by
\begin{equation}\label{fEM3}
\overline{{\bf f}_{21}}=-\overline{{\bf f}_{12}}=-\frac{T_0}{L}\kk.
\end{equation}
The net force density on the ME region is thus 
\begin{equation}
{\bf f}_{\rm vac} 
=\overline{{\bf f}_{12}}+\overline{{\bf f}_{21}}={\bf 0},
\end{equation}
i.e. the net force density on the ME fluid is zero for steady ME susceptibility (steady fields): the first law is safe! This agrees with the prediction of Obukhov \& Hehl for a solid ME slab \cite{ObukhovHehl08} and has the same physical origin. That the net force on the ME medium should be zero can also be seen in a physically more immediate way by a consideration of kinetic argument vacuum modes \cite{Bongs12}. Right (left) travelling modes gain (lose) momentum on entry across the  21 (12) surface, but lose (gain) equal and opposite monentum on exiting through the 12 (21) surface. The net momentum transfer is zero.
\begin{figure}[tbph]
\centering
\includegraphics[width=0.5\linewidth]{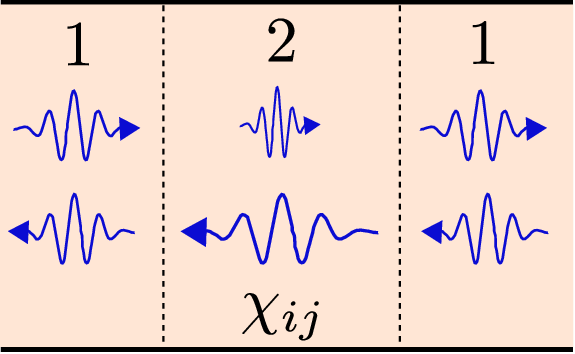}
\caption{The wavelength of virtual photons entering the surface 21 (12) shortens (lengthens) due to ME response and momentum is transferred to the fluid, but it lengthens (shortens) once more at exit by the 12 (21) surface. Thus the net momentum transfer by the vacuum on the magnetoelectric is zero: there cannot be a steady Feigel effect.}\label{modemomtransfer}
\end{figure}

I have shown that a steady Feigel effect is incompatible with the first law of thermodynamics and QED. Revising my analysis in \cite{Croze12} to properly account for boundaries shows that no net transfer of momentum between the vacuum and magnetoelectrics with steady susceptibilities is possible. Interestingly, Obukhov \& Hehl \cite{ObukhovHehl08} claim that a classical steady Feigel effect (a ME slab placed in counter propagating laser beams) might instead exist. However, the analysis in this paper, easily applied to optical EM modes, implies that a steady Feigel effect should not be observed for light either. Of course here considered infinite wave trains (as do \cite{ObukhovHehl08}). For pulses, an optical Feigel effect seems more feasible, but will not be discussed fur here here.%
\begin{figure}[tbph]
\centering
\includegraphics[width=0.6\linewidth]{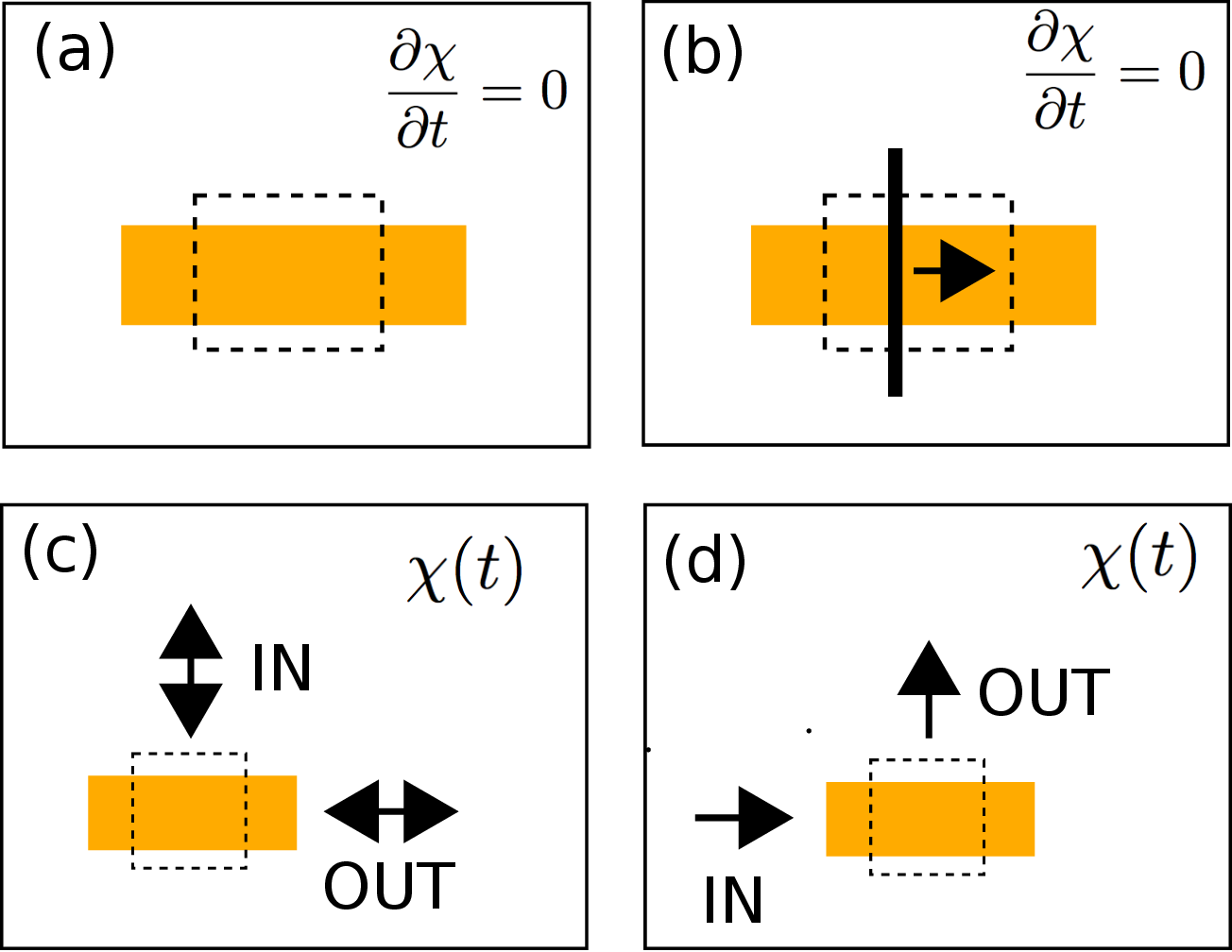}
\caption{Feigel effects: (a) the bounded steady Feigel effect, no energy flows; (b) single plate Casimir-Feigel effect. Energy is transferred to a plate, which drifts to the boundary of the ME region. An equal amount of work needs to performed to restore the plate to its initial position; (c) unsteady Feigel effect: the energy of the fields is transferred to the ME medium via the quantum vacuum (but is much smaller than the classical effect); (d) in a dynamical Feigel effect of an accelerated ME medium. The energy input released as light}\label{modemomtransfer}
\end{figure}

While the steady Feigel effect is not possible, variations may be compatible with conservation of energy. These are summarised in figure (\ref{modemomtransfer}). An intriguing possibility is a single-plate Casimir-Feigel effect: a plate is submerged in an ME liquid with its surface perpendicular to the ME optical axis. The asymmetric vacuum stress on the plate surfaces should cause it to drift across the region. In this Casimir-Feigel one has to do work to restore the plate to its original position, so any cyclic extraction of energy from the vacuum is not possible. A theoretical analysis of this effect is left to future work. It will be interesting to see if the predicted drift speed of the plate is sizeable enough to be measured. When the susceptibility of magnetoelectrics is time dependent (e.g. when unsteady fields act on ME fluids), the Feigel effect should also be possible. Estimates on the magnitude of this transient compared to the classical effects due to polarisation currents effect reveal it might be too stall to measure using current experimental techniques \cite{Forgan12}. Instead of oscillating fields to obtain a time-dependent response, one could instead mechanically accelerate the ME material. This would be a new dynamical Feigel effect, analogous to the Casimir counterpart recently verified experimentally in elegant experiments \cite{Lahteenmakiaetal13}.

To conclude, I hoped the analysis in this paper will inspire theoreticians to formulate testable predictions for the Feigel effect and its variations. Of course lack of testable theories should not prevent experimentalists from searching for Feigel-like effects in magnetoelectrics and other anisotropic or chiral molecules and media.

\begin{acknowledgments}
I acknowledge discussions with A. Holmes, E. Forgan, K. Bongs, M. Cates, O. Birkeland and I. Brevik. I am grateful to S. Paraoanu for suggesting the possibility of a dynamical Feigel effect. I acknowledge funding from EPSRC (EP/J004847/1) and the Winton Programme for the Physics of Sustainability.
\end{acknowledgments}

\bibliography{CrozeFeigelPRARapid}



\end{document}